\documentclass[a4paper]{article}
\usepackage{INTERSPEECH2022}
\usepackage{amsmath,graphicx}
\usepackage{cite,subfigure,amssymb,tablefootnote,multirow}
\usepackage{tipa}

\usepackage[colorlinks,
            linkcolor=red,
            anchorcolor=blue,
            citecolor=green
            ]{hyperref}

\title{SCaLa: Supervised Contrastive Learning for End-to-End Speech Recognition}
\name{Li Fu, Xiaoxiao Li, Runyu Wang, Lu Fan, Zhengchen Zhang,\\Meng Chen, Youzheng Wu, Xiaodong He}
\address{JD AI Research, Beijing, China}
\email{\{fuli3,lixiaoxiao10,wangrunyu3,fanlu,zhangzhengchen1,
chenmeng20,wuyouzheng1,hexiaodong\}@jd.com}

\begin{document}

\maketitle
\begin{abstract}
End-to-end Automatic Speech Recognition (ASR) models are usually trained to optimize the loss of the whole token sequence, while neglecting explicit phonemic-granularity supervision. This could result in recognition errors due to similar-phoneme confusion or phoneme reduction. To alleviate this problem, we propose a novel framework based on \underline{S}upervised \underline{C}ontr\underline{a}stive \underline{L}e\underline{a}rning (SCaLa) to enhance phonemic representation learning for end-to-end ASR systems. Specifically, we extend the self-supervised Masked Contrastive Predictive Coding (MCPC) to a fully-supervised setting, where the supervision is applied in the following way. First, SCaLa masks variable-length encoder features according to phoneme boundaries given phoneme forced-alignment extracted from a pre-trained acoustic model; it then predicts the masked features via contrastive learning. The forced-alignment can provide phoneme labels to mitigate the noise introduced by positive-negative pairs in self-supervised MCPC. Experiments on reading and spontaneous speech datasets show that our proposed approach achieves 2.8 and 1.4 points Character Error Rate (CER) absolute reductions compared to the baseline, respectively.
\end{abstract}

\section{Introduction}
In recent years, the accuracy of end-to-end Automatic Speech Recognition (ASR) systems has been significantly improved for various datasets~\cite{Chan_2016_ICASSP,Bahdanau_2016_ICASSP,Prabhavalkar_2017_Interspeech,Li_2020_Interspeech,Li_2021_recent}. Typically, the models are optimized to improve the average performance over the entire sequence when mapping an input speech to an output character or word sequence, lacking explicit phoneme level supervision. They are powerful enough to learn latent representation partially corresponding to phonemes from each frame~\cite{Belinkov_2017_NeurIPS}. However, the models are still not robust to phonemic issues like similar-phoneme confusion~\cite{Fang_2020_SIGIR} as well as consonant or vowel reduction~\cite{Ma_2021_Interspeech}.

\begin{figure}[t]
	\centering
	\subfigure[Phoneme-level forced-alignment of a Mandarin speech]{
		\begin{minipage}[t]{1\linewidth}
			\centering
			\includegraphics[width=0.95\linewidth]{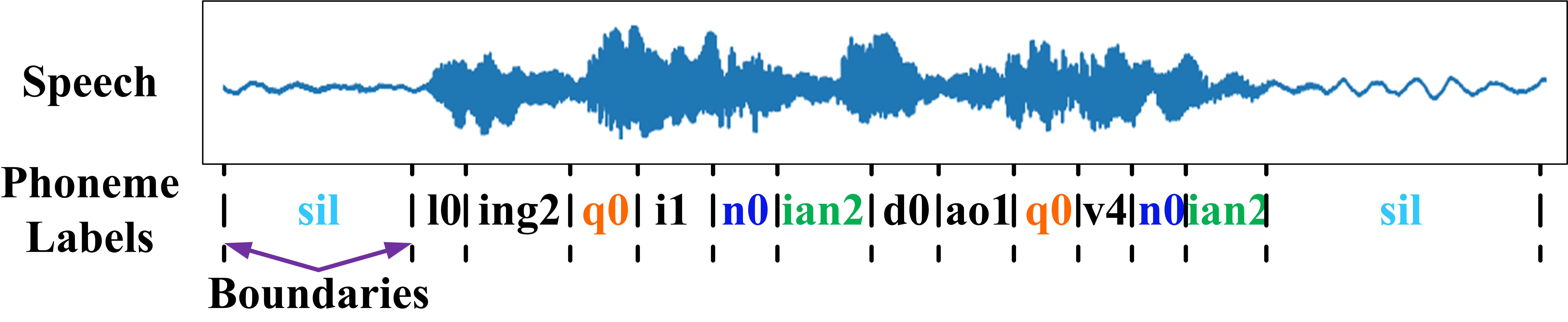}
			\label{fig1_a}
		\end{minipage}
	}
	\hfill
	\subfigure[Unlike self-supervised MCPC (in purple/regular font), SCaLa (in red/italic font) masks phonemes based on boundary information on encoder features$^1$, and constructs contrastive feature pairs from context features and target features with mitigation of noisy negative pairs. If the first ``q0" in context features is selected, the second ``q0" in target features should not be selected as the negative features for contrasting.]{
		\begin{minipage}[t]{1\linewidth}
			\centering
			\includegraphics[width=0.95\linewidth]{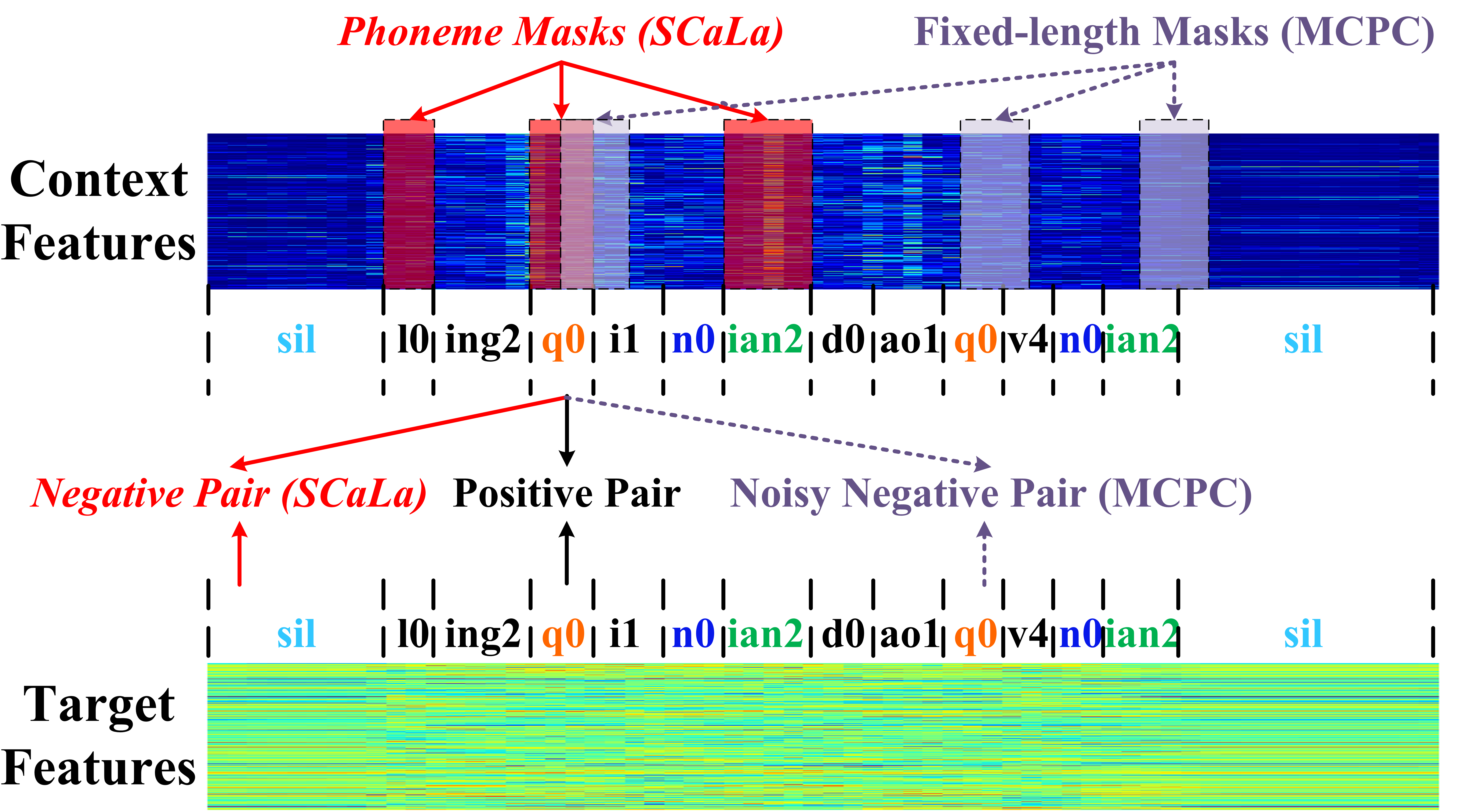}
			\label{fig1_b}
		\end{minipage}
	}%
	\centering
	\caption{Advantages of SCaLa using phoneme-level forced-alignment (involves labels and boundaries) as supervision over MCPC~\cite{Baevski_2020_NeurIPS}, in terms of masking and contrasting strategies.
	}
	\label{fig1}
\vspace{-0.1in}
\end{figure}

Contrastive learning has shown great potential in addressing the phonemic issues of ASR tasks -- a variety of masking and contrasting strategies have been proposed to learn speech representations for downstream tasks~\cite{Baevski_2020_ICASSP,Ravanelli_2020_ICASSP,Liu_2020_ICASSP,Hsu_2021_TASLP,Pasad_2021_ASRU}. Recently, most existing contrastive learning based ASR systems assume a self-supervised setting~\cite{Liu_2021_survey}. Masked Contrastive Predictive Coding (MCPC), proposed in Wav2vec2.0~\cite{Baevski_2020_NeurIPS}, is one of the most representative methods. It masks a consecutive frame of the encoder features with a fixed/random length, then selects anchor and positive samples to obtain positive pairs from the same masked indices of the defined context features and target features, and randomly selects the negative samples from other indices of the target features to obtain negative pairs. The model is then trained to discriminate the anchor/positive features from a set of negative features via a contrastive task. However, applying MCPC to unlabeled data is challenging in mask length selection and noise reduction (caused by negative samples). As shown in Fig. \ref{fig1}: (1) Since phonemes usually have various lengths in speech, masking with a fixed/random length would ignore the boundaries between adjacent phonemes, which may damage the model on learning phonemic representation effectively~\cite{Ma_2021_Interspeech,Wang_2019_Interspeech}. (2) In contrastive learning, the indices of negative features for contrasting are randomly selected~\cite{Baevski_2020_NeurIPS}. Hence, there might exist noisy negative pairs. For example, the anchor/positive-negative pairs may come from the same phoneme, or both of them may be silence or background noise, etc. As referred in~\cite{Khosla_2020_NeurIPS}, noisy negative samples will compromise the effectiveness of the feature representation. 

\footnotetext[1]{Masking is for selecting contrastive pairs and is actually performed on encoded features only; the masks on the context features are presented here merely for showing the features being selected.}

\begin{figure*}[t]
  \centering
  \includegraphics[width=0.62\textwidth]{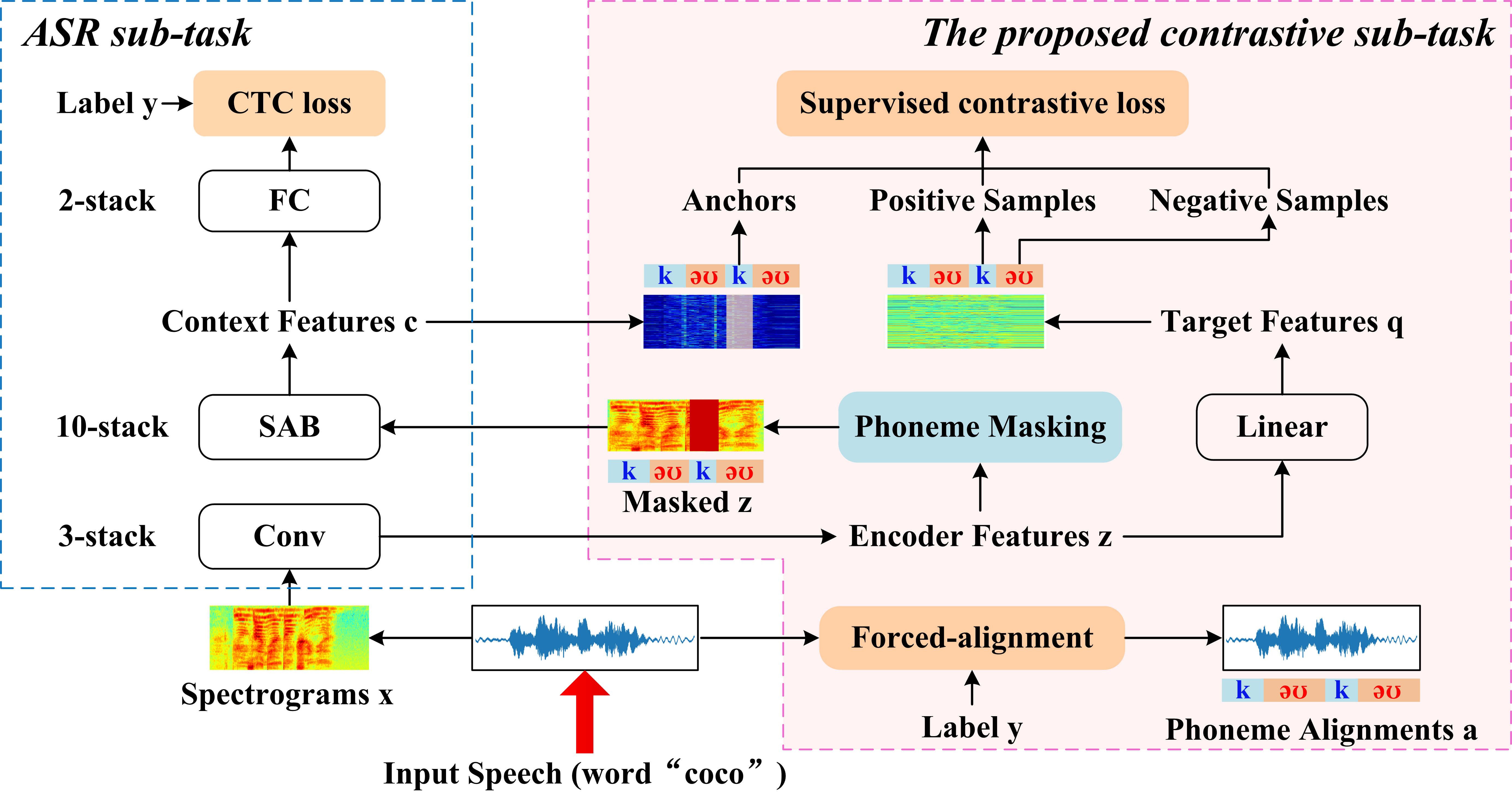}
  \caption{Model architecture of SCaLa. A CTC-based ASR sub-task is combined with the proposed contrastive sub-task that leverages a forced-alignment model to perform phoneme masking and contrastive learning. The backbone CTC-base ASR network is composed of a successive stack of Convolutional (Conv) layers, Self-Attention Blocks (SABs), and Fully Connected (FC) layers. In the proposed contrastive sub-task, masked items from context features are selected as anchors (associated with phoneme ``k"), with the same indices in target features taken as positive samples correspondingly; while items with other indices from target features are selected as negative samples (associated with different phonemes, e.g. ``\textschwa\textupsilon").}
  \label{fig2}
\end{figure*}

To address the above challenges in self-supervised MCPC, we propose a novel framework named \underline{S}upervised \underline{C}ontr\underline{a}stive \underline{L}e\underline{a}rning (SCaLa) for end-to-end ASR systems. Unlike previous self-supervised studies, SCaLa applies MCPC in a fully-supervised manner, which has the following two advantages in masking and contrasting strategies, respectively. First, the mask length is customized for each particular phoneme duration, i.e., the masked unit can be a complete phoneme. This will help improve the prediction accuracy of reduced consonants or vowels in speech~\cite{Ma_2021_Interspeech}. Specifically, we perform forced-alignment between utterances and their labeled transcription to obtain the phoneme labels and boundaries (shown in Fig. \ref{fig1_a}). Then the encoder features are masked according to phoneme boundaries to help the model learn phonemic representation explicitly. Second, the noisy anchor/positive-negative pairs are mitigated by selecting the negative features based on the phoneme forced-alignment labels (shown in Fig. \ref{fig1_b}). Although the alignments may not be perfect, the proposed method can still largely reduce noisy negative pairs empirically, e.g., same-phoneme pairs or silence pairs. Hence the ASR model can learn better latent representation by phoneme discrimination. 

To train the ASR model, our SCaLa involves two sub-tasks: (1) an ASR sub-task to directly generate character or word sequences from acoustic features, and (2) a contrastive sub-task to predict masked phonemes to improve phoneme discrimination via contrastive learning. We combine the two sub-tasks to help the model learn the representations of character or word sequences and phoneme level information at the same time, to improve the performance on speech recognition tasks. Our main contributions are: 1) To the best of our knowledge, this is the first work extending the self-supervised MCPC approach to fully-supervised ASR systems; 2) We propose a framework named SCaLa for end-to-end ASR to enhance phoneme-level representation learning; 3) We show the effectiveness of our method, with discussion, on both reading and spontaneous speech data.

The remainder of this paper is organized as follows. Section 2 presents the related work. Section 3 is the details of the proposed method. Section 4 shows the experimental results. Finally, the conclusions and future work are given in Section 5.

\section{Related Work}
\label{sec:relatedwork}
Contrastive learning based on Predictive Coding (PC) has been widely used in self-supervised ASR training. Contrastive PC (CPC)~\cite{Oord_2018_arXiv} and its variants~\cite{Chung_2019_Interspeech} predict future speech segments based on the past ones for pre-training unidirectional networks. The CPC-based method was adopted for ASR tasks in Wav2vec~\cite{Schneider_2019_Interspeech} and Vq-wav2vec to learn discrete representation of speech units~\cite{Baevski_2020_ICLR}. Masked PC (MPC) was proposed in~\cite{Jiang_2019_arXiv}, which can improve the performance of Transformer based ASR systems by predicting masked encoder features. In Wav2vec2.0~\cite{Baevski_2020_NeurIPS}, the authors proposed MCPC, which involved contrastive learning on top of MPC, with significant gain in performance on downstream ASR tasks. However, as mentioned in~\cite{Wang_2021_ICML}, the self-supervised paradigm of Wav2vec2.0 needs to be carefully designed, and the representation is difficult to interpret. To help the model learn a more meaningful speech representation with MCPC, UniSpeech~\cite{Wang_2021_ICML} and JUST~\cite{Bai_2022_ICASSP} were proposed by combining labeled and unlabeled speech for training in a multitask manner; however, labels were still not exploited for MCPC performing. To apply contrastive learning for accented 
ASR, the authors of~\cite{Han_2021_arXiv} adopted SimCLR~\cite{Chen_2020_ICML} in the computer vision domain, and then generated contrastive positive pairs from the model's output corresponding to letter-level tokens using various data augmentation methods. Differently, our SCaLa uses forced-alignment results to build contrastive pairs, which ensures these contrastive pairs to be independent from the model training. Masking strategy based on confidence has been introduced into self-supervised training~\cite{Baskar_2022_arXiv}. Besides the supervised setting, our SCaLa concerns more about generating contrastive pairs by masking phonemic-level encoder features to learn phonemic latent representations explicitly.

\section{Proposed Method}
\label{sec:method}
\subsection{Model architecture}
\label{subsec:model}
The model architecture of SCaLa is shown in Fig.~\ref{fig2}, which consists of an ASR sub-task and the proposed contrastive sub-task.

As for the ASR sub-task, the representative Connectionist Temporal Classification (CTC) framework used in ASR training~\cite{Schneider_2019_Interspeech,Baevski_2020_NeurIPS,Wang_2021_ICML} is adopted as our backbone model. Our experiments use the typical model network, which is composed of a successive stack of Convolutional (Conv) layers, Self-Attention Blocks (SABs), and Fully Connected (FC) layers~\cite{FU_2021_arXiv}. Given a sequence of ${d_s}$-dimensional acoustic spectrograms ${\mathbf{x}\in\mathbf{R}^{{d_s}\times{T}}}$ with length ${T}$, the ASR model tries to predict the labeled character sequence $\mathbf{y}\in\mathbf{L}^{N}$ with length ${N}$, where $\mathbf{L}$ is the size of the finite label character. The output of the last Conv layer is denoted as encoder features ${\mathbf{z}\in\mathbf{R}^{{d_f}\times{S}}}$, where ${d_f}$ is the latent feature dimension, and ${S}$ is the sequence length. Note that ${S}$ is less than ${T}$ after subsampling by the Conv layers~\cite{Amodei_2016_ICML}. The phoneme-level forced-alignment results are denoted by ${\mathbf{a}\in\mathbf{R}^{{S}}}$~\cite{Gorman_2011_CA}. Merely for brevity, we assume that each item in $\mathbf{a}$ is a phoneme label of items in the encoder features $\mathbf{z}$ instead of the input speech. The items in $\mathbf{z}$ are masked with some prescribed probability during training. The masked features are fed into SABs that yields context features ${\mathbf{c}\in\mathbf{R}^{{d_f}\times{S}}}$. 

Regarding the proposed contrastive sub-task, 
a linear layer is adopted like in~\cite{Zhang_2020_NeurIPS,Bai_2021_arixv} to obtain the contrastive target features ${\mathbf{q}\in\mathbf{R}^{{d_f}\times{S}}}$. 
Then we select masked items from context features as the anchors~\cite{Baevski_2020_NeurIPS}. Items in target features with the same indices are taken as the positive samples, while items in target features with indices from other alignment labels are taken as negative samples. For the contrastive sub-task, a contrastive loss is involved (in our loss function described next) to guide the model to decrease the similarities between the anchor-negative pairs, and to increase the similarities between the anchor-positive pairs.

\subsection{Loss functions}
\label{subsec:lossfunction}
We combine the ASR sub-task and the contrastive sub-task to help the model learn the representation of character sequences and phoneme-level features simultaneously, and to improve the performance on speech recognition tasks. The two losses of SCaLa corresponding to the two sub-tasks are (1) a CTC loss based on phoneme masking ${{L}_{\rm CTC}}$ and (2) a supervised contrastive loss with phonemic-granularity supervision ${{L}_{\rm SCL}}$.

\subsubsection{CTC loss based on phoneme masking}
\label{sssec:ctcloss}
Phoneme labels and boundaries are used for masking to help the model enhance phonemic representation~\cite{Ma_2021_Interspeech}. Specifically, a HMM-DNN acoustic model is trained offline using Kaldi~\cite{Povey_2011_ASRU} to get the phoneme-level label of each frame and the corresponding boundaries. An example is shown in Fig.~\ref{fig1_a}. During the training, we randomly sample a set of start indices of encoder features with a certain probability $p_e$ (We set $p_e=6.5\%$ like in~\cite{Baevski_2020_NeurIPS}). A total of $P$ phonemes adjacent to the start indices are integrally masked by leveraging the phoneme boundaries given forced-alignment. The choice of $P$ is experimentally studied in Sec. \ref{subsec:main_} (see Fig. \ref{fig3}). For a data-label pair \{${\mathbf {x,y}}$\} and the forced-alignment result $\mathbf{a}$, the CTC loss based on phoneme masking is obtained as 

\begin{equation}
  {{L}_{\rm CTC}}=-{\log}\sum_{\boldsymbol{\pi}\in {\boldsymbol \phi({\mathbf {x,y}}) }}{p}({\boldsymbol{\pi}|\mathbf{x}},{\mathbf{a}},p_e)
  \label{eq1}
\end{equation}
where a valid CTC path $\boldsymbol{\pi}$ is a variant of the transcription $\mathbf{y}$ that allows occurrences of blank tokens and repetitions. The set $\boldsymbol \phi({\mathbf {x,y}})$ includes all valid CTC paths~\cite{Graves_2006_ICML}.

\subsubsection{Contrastive loss with phonemic-granularity supervision}
\label{sssec:Contrastive loss}
Supervised contrastive learning~\cite{Khosla_2020_NeurIPS} aims to improve the robustness of feature representation by discriminating an anchor/positive phoneme from a set of negative phonemes. In particular, to reduce noisy negative pairs, features having the same phoneme label with the masked phoneme are avoided from the negative phoneme sets. As shown in Fig.~\ref{fig1_b}, the second ``q0" of target features will not be selected as negative phonemes to be contrasted with the first ``q0" of context features. Accordingly, the supervised contrastive loss is defined as
\begin{equation}
    {{L}_{\rm SCL}}=-\frac{1}{\left |\boldsymbol M \right |}\sum_{{m}\in \boldsymbol M}{\log}\frac{e^{sim(\mathbf c_m,\mathbf q_m)/\tau}}{\sum_{n\in \boldsymbol{N_m}}e^{sim(\mathbf c_m,\mathbf q_{n})/\tau}}
    \label{eq2}
\end{equation}
where $\boldsymbol M$ is the set of all masked indices of encoder features, and ${\left |\boldsymbol M \right |}$ is the number of masked indices; $\mathbf c_m$ and $\mathbf q_m$ are the $m^{th}$ vectors in context features $\mathbf c$ and target features $\mathbf q$, respectively; $sim({\boldsymbol \alpha},{\boldsymbol \beta})={\boldsymbol \alpha}^T {\boldsymbol \beta}/({\left | \right |} {\boldsymbol \alpha}{\left | \right |}\ {\left | \right |} {\boldsymbol \beta}{\left | \right |})$ is the cosine similarity; $\tau$ is a temperature scale. The index set $\boldsymbol {N_m}$ consists of the masked index $m$ and a negative index set $\boldsymbol K$, which are uniformly sampled from all indices except those having the same alignment label with the masked phoneme $\mathbf a_m$, i.e. $\mathbf a_{k} \neq \mathbf a_m,
~\forall {k} \in \boldsymbol K$. We set $\tau=0.1$ and the number of negative indices ${\left |\boldsymbol K \right |}=100$ in our experiments -- the same as~\cite{Baevski_2020_NeurIPS}.

\subsection{Model training}
\label{subsec:modeltraining}
Following~\cite{Talnikar_2021_ICASSP}, an alternate minimization training method is employed in our proposed method as well. The training losses ${L}_{\rm CTC}$ and ${{L}_{\rm SCL}}$ are minimized alternately with a balanced ratio, i.e. 1:1, to update the model parameters. The main advantage of alternating training is that the learning rates of ASR sub-task optimizer and contrastive sub-task optimizer are separated~\cite{Talnikar_2021_ICASSP}. In our experiments, we find that a single optimizer resulting from a weighted sum of the two loss functions would result in a slightly higher CER, and the ratio does not significantly influence the performance of our method.

\section{Experiments and Discussion}
\label{sec:results}
\subsection{Experimental setup}
\label{subsec:exp_setup}
Two datasets with different speaking styles are used in our experiments: 1) reading speech data: the open-source Aishell-1 which contains 170 hours of Mandarin speech with 16kHz sampling rate~\cite{Bu_2017_COCOSDA}; and 2) spontaneous speech data: an in-house Mandarin conversational Telephony (JD-Tel) dataset which contains 1500 hours of speech with 8kHz sampling rate. For Aishell-1, the original train-test split is used. For the other one, 10$\%$ of the samples are randomly selected for testing.

The 80-dimensional Mel-spectrograms are used as the input to the network. The frame size and step size are 20ms and 10ms, respectively. Our model contains 3 Conv layers, 10 SABs, and 2 FCs, as shown in Fig.~\ref{fig2}. Please refer to~\cite{FU_2021_arXiv} for more details about the model.

\begin{table}[b]
    \vspace{-0.1in}
	\centering  
	\caption{System performance in CER ($\%$).}
	\label{table1}  
	\resizebox{0.95\linewidth}{!}{
	\begin{tabular}{l|c|c}  
		\hline  
		{\bf Testing data} & {\bf Aishell-1} & {\bf JD-Tel} \\
		\hline
		{\bf Speaking style} & {\bf reading} & {\bf spontaneous}\\
		\hline
	    \hline
		Chain model~\cite{Yu_2021_arXiv} & 7.5 & 15.9\\
		Wav2vec2.0~\cite{Baevski_2020_NeurIPS,Yuan_2021_arXiv}&5.3&16.3\\
		WeNet CTC-conformer\cite{Yao_2021_Interspeech}&&\\
		\ \ \ \ \ \ \ w/ CTC prefix beam search& 5.9 & 15.6\\
		\ \ \ \ \ \ \ w/ attention rescoring&5.3&14.7\\
		\hline
		CTC (Baseline)\cite{FU_2021_arXiv}&6.7&15.3\\
		CTC$+$phoneme mask\cite{Ma_2021_Interspeech}&5.1&14.8\\
		SCaLa & {\bf 3.9}&{\bf 13.9}\\
        \hline
	\end{tabular}
	}
\end{table}

Our models are trained on 4 NVIDIA V100 GPUs with mini-batch size 128. For the alternate loss minimization in Sec. \ref{subsec:modeltraining}, the Learning Rate (LR) for ${L}_{\rm CTC}$ is $2.5 \times 10^{-5}$, and $5\times 10^{-4}$ for ${{L}_{\rm SCL}}$. The LR of ${L}_{\rm CTC}$ is unchanged, while the LR of ${{L}_{\rm SCL}}$ is decayed to $5\times 10^{-5}$ ultimately. 

\subsection{Main experiment and ablation study}
\label{subsec:main_}
As shown in Table~\ref{table1}, we compare SCaLa with state-of-the-art ASR systems including hybrid (i.e. the chain model, a type of improved DNN-HMM)~\cite{Yu_2021_arXiv}, end-to-end~\cite{Yao_2021_Interspeech,FU_2021_arXiv,Ma_2021_Interspeech}, and self-supervised learning~\cite{Baevski_2020_NeurIPS,Yuan_2021_arXiv}. Experimental results show that existing end-to-end systems~\cite{Ma_2021_Interspeech,FU_2021_arXiv,Yao_2021_Interspeech} and self-supervised learning~\cite{Baevski_2020_NeurIPS,Yuan_2021_arXiv} largely outperform the chain model that based on forced-alignment~\cite{Yu_2021_arXiv}. The performance of CTC models~\cite{FU_2021_arXiv} can also benefit from phoneme mask strategies~\cite{Ma_2021_Interspeech}. Compared with the existing methods, our SCaLa achieves the best performance. Numerically, it outperforms the traditional CTC models~\cite{FU_2021_arXiv} with 2.8 and 1.4 points CER absolute reductions on reading and spontaneous speech data, respectively. 


\begin{figure}
\vspace{-0.2in}
\centering
\subfigure[]{\label{fig:subfig:a}
\includegraphics[width=0.47\linewidth]{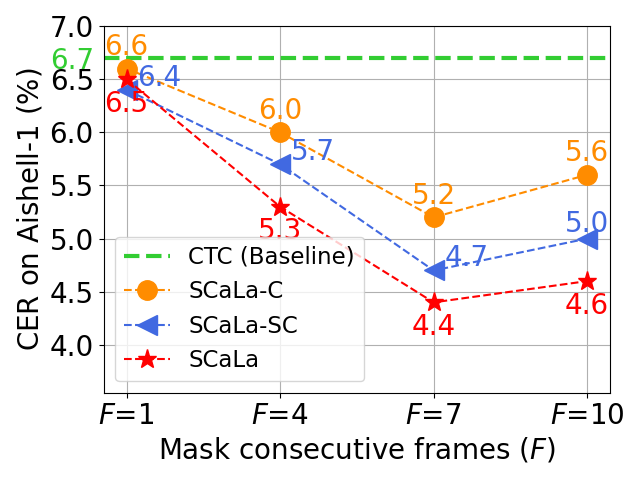}}
\subfigure[]{\label{fig:subfig:b}
\includegraphics[width=0.47\linewidth]{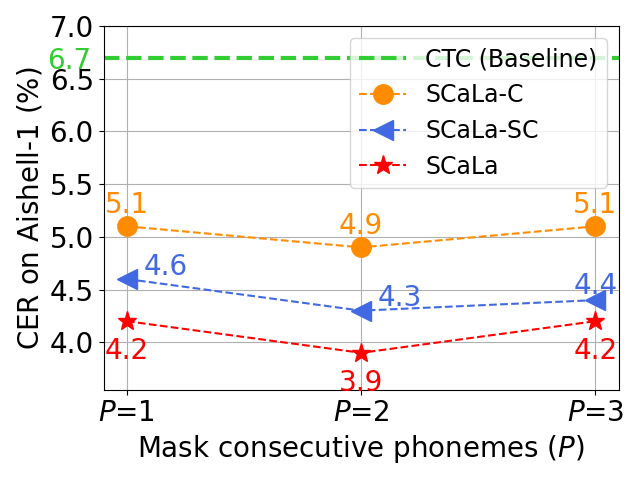}}
\vfill
\subfigure[]{\label{fig:subfig:a}
\includegraphics[width=0.47\linewidth]{{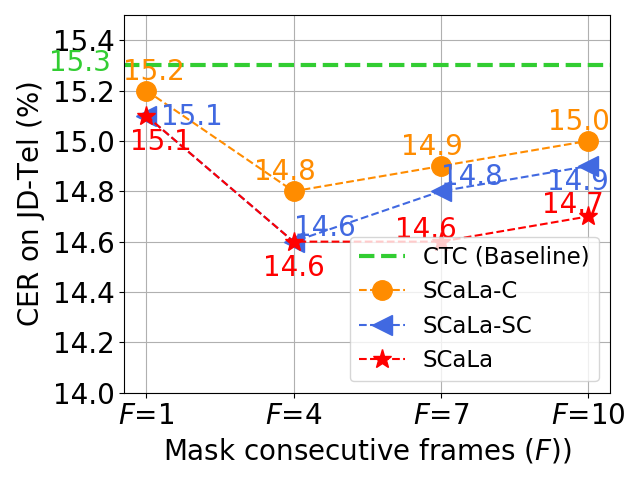}}}
\subfigure[]{\label{fig:subfig:b}
\includegraphics[width=0.47\linewidth]{{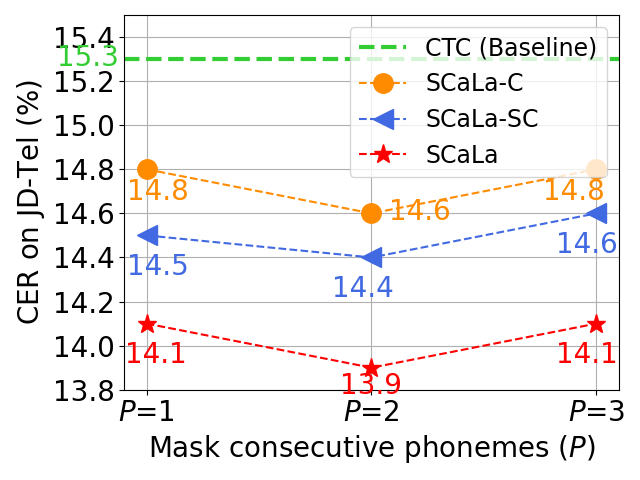}}}
\caption{Ablation study where CERs for SCaLa compared with: 1) SCaLa using random samplings for the contrastive sub-task without supervision (SCaLa-SC), and 2) SCaLa without the contrastive sub-task (SCaLa-C), for a variety of mask settings on reading and spontaneous speech data. The performance of the traditional CTC method (Baseline) is included for reference.}
\label{fig3}
\vspace{-0.15in}
\end{figure}

As ablation studies on phoneme masking and contrastive learning: (1) We compare two masking methods: masking $F$ consecutive frames (fixed-length masks in~\cite{Baevski_2020_NeurIPS}) and masking $P$ consecutive phonemes (phoneme masks in SCaLa) where $F\in\{1,4,7,10\}$ and $P\in\{1,2,3\}$. In the experiments, the forced-alignment results show that the average phoneme lengths are 3.3 and 2.6 frames for the reading and spontaneous speech data, respectively. (2) Different contrastive strategies including the proposed SCaLa, SCaLa using random samplings for the contrastive sub-task without supervision (SCaLa-SC), and SCaLa without the contrastive sub-task (SCaLa-C) are also compared. The results for these ablation studies are shown in Fig.~\ref{fig3}. The CERs curves of different methods indicate that supervised contrastive learning can significantly reduce CERs. Moreover, the results of different masking strategies on both sides of the figures show that phoneme masking outperforms the fixed-length masking in ASR tasks. The results also indicate that our SCaLa achieved the best performance when $P=2$. We infer that the optimal value of $P$ is related to the fact that a character in Mandarin usually contains two phonemes~\cite{Huang_2019_ASRU}. Larger mask length creates a heavy burden on the model learning because there is an overly large variety in the masked contents, while smaller ones influence the training very little~\cite{Baevski_2020_NeurIPS}.

\subsection{Analysis of SCaLa}
\label{subsec:Deeper_insight}
To further evaluate the effectiveness of SCaLa, we analyze the proposed method from the following three perspectives:

\noindent{\bf Robustness to phonemic issues.} 
The phonemic issues, such as similar-phoneme confusion and phoneme-reduction usually introduce substitution and deletion errors~\cite{Ma_2021_Interspeech}. The detailed CER, including substitution (SUB), deletion (DEL) and insert (INS) error rates, of SCaLa and the baseline CTC method are shown in Table~\ref{table2}. The results show that SCaLa achieves significant reductions on substitution and deletion errors. We also observe that the performance improvement of SCaLa on reading speech data is more than that on spontaneous speech data. The reasons may be that recognition tasks on spontaneous speech are more challenging, and that the given forced-alignment results may not be accurate enough. Nevertheless, our method can still improve the performance of speech recognition.

\begin{figure}[t]
  \vspace{-0.15in}
  \centering
  \includegraphics[width=0.4\textwidth]{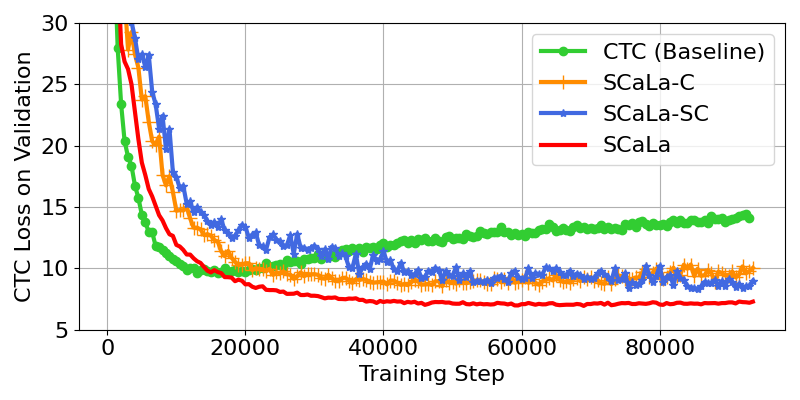}
  \caption{Regularization effect of SCaLa with $P=2$ on Aishell-1 (Each training epoch contains 936 steps in the experiments).}
  \label{fig4}
 \vspace{-0.1in}
\end{figure}

\begin{table}[ht]
	\centering  
	\caption{Detailed CER ($\%$), including substitution (SUB), deletion (DEL) and insert (INS) error rates, for SCaLa and the baseline CTC method on reading (Aishell-1) and spontaneous (JD-Tel) speech data.}
	\label{table2}  
	\resizebox{0.99\linewidth}{!}{
	\begin{tabular}{c|c||c|c|c||c}  
		\hline  
		\textbf{Testing data}& \textbf{Methods}
        & \textbf{SUB} & \textbf{DEL} & \textbf{INS} &\textbf{CER}\\
		\hline
		\hline
		\multirow{3}{*}{Aishell-1}&CTC (Baseline)~\cite{FU_2021_arXiv} &5.2 & 1.4 & {\bf 0.1}& 6.7 \\
		&SCaLa-SC&3.8&0.4&{\bf 0.1}&4.3\\
		&SCaLa & {\bf 3.5} & {\bf 0.3} & {\bf 0.1}& {\bf 3.9} \\
        \hline
        \multirow{3}{*}{JD-Tel}&CTC (Baseline)~\cite{FU_2021_arXiv} &10.2 & 4.7 & 0.4& 15.3\\
	    &SCaLa-SC&9.9&4.2&{\bf 0.3}&14.4\\
		&SCaLa & {\bf 9.7} & {\bf 3.9} & {\bf 0.3}& {\bf 13.9} \\
        \hline
	\end{tabular}
	}
	\vspace{-0.1in}
\end{table}

\noindent{\bf Noisy negative reduction.} We use the forced-alignment results to count the proportion of noisy negative pairs among all the negative pairs for contrasting. We find that the noisy negative rates of self-supervised MCPC are 10.21$\%$ and 14.60$\%$ for the two datasets, respectively. The number is non-negligible~\cite{Khosla_2020_NeurIPS}. As shown in Table~{\ref{table2}}, SCaLa improves system performance compared to SCaLa-SC by reducing the noisy negatives.

\noindent{\bf Regularization effect.} Fig.~\ref{fig4} shows the CTC losses of SCaLa, SCaLa-SC, SCaLa-C and the baseline CTC method on the validation data of Aishell-1. SCaLa obtains the lowest and smoothest loss curve compared with the other three methods, which indicates the regularization effect to the training.

\section{Conclusion}
In this paper, a novel framework named SCaLa has been proposed for ASR training. It extends the self-supervised MCPC approach to a fully-supervised setting. The labels are effectively leveraged to enhance ASR models to learn phoneme representation. SCaLa significantly improved the performance on both reading and spontaneous Mandarin speech data compared to the baseline methods. In the future work, the performance on more other masking strategies and languages will be evaluated.


\end{document}